# Comment on

**"Water sources and kidney function: investigating chronic kidney disease of unknown etiology in a prospective study",**
**NPJ Clean Water 4, 50, (2021)**
**by P. Vlahos *et al*.**
https://doi.org/10.1038/s41545-021-00141-2

by

M. W. C. Dharma-wardana

National Research Council of Canada, Ottawa, Canada K1A 0R6
and
Université de Montreal, Montreal, Canada.  H3C 3J7

Funding: No funding sources

Competing Interests: None

email:
chandre.dharma-wardana@nrc-cnrc.gc.ca
chandre.dharma@yahoo.ca

**Vlahos *et al*., Ref. 1, NPJ Clean water. 4,** 50 (2021) **have reported the presence of pesticide contamination above safe levels** in a "single time-point analysis" of well water in a region in Sri Lanka where chronic kidney disease of unknown etiology (CKDu) is endemic. They conclude "that agrochemical use in paddy and other agricultural practices … of the Green Revolution in Sri Lanka may now be contributing to ill health, rapid progression of disease, and mortality". The authors also propose "reducing … agrochemical contaminants in Sri Lanka and other tropical countries to reduce … CKDu. These conclusions, based on what they call a "single time-point analysis", tantamount to an identification of the etiology of CKDu are unsupported by the evidence presented by Vlahos *et al*. They do not satisfy, say, even the simplest of Bradford-Hill criteria for causation. In particular, (i) similar but non-persistent pesticide excesses have been detected *sporadically* in most parts of the country including where there is no CKDu; (ii) the pesticides reported in (1) cause both hepatotoxicity *and* glomerular damage while CKDu is associated with tubulo- interstitial damage where no hepatotoxic symptoms have been reported; (iii) the pesticides detected have short half-lives and are used over short periods during farming; so the one time-point analysis is inadequate and misleading; (iv) farming communities that use pesticides in the same way but remain essentially *without* CKDu are found to exist adjacent to communities *with* CKDu; (v) the CKDu prevalence seems to correlate with local geomorphology but without correlation to agriculture which is practiced in most parts of the country.

A major theme of Vlahos study is the negative effects of the Green Revolution. It sharply increased food production, eliminated malnutrition, nearly doubled life expectancy, and virtually eliminated infant mortality, but also increased environmental pollution. The increased life expectancy has led to the expected proportionate increase in non-communicable disease, with no strong non-linear effects. However, the rise of CKDu forces one to review the situation. Vlahos *et al.* in their study once again raise the important issue of pesticides used in the Green revolution and their possible role in CKDu.

When CKDu was detected in Sri Lanka, Mesoamerica, India etc. (2), an agrochemical etiology was a prime suspect. The favoured suspects were traces of heavy-metals like As, Cd etc. found in agrochemicals. Strong claims were made against glyphosate and arsenic, leading to a ban on glyphosate use, although the claim remains unsubstantiated. Already in 2012 it was evident that heavy metal concentrations in the water and soil in the affected regions were negligible (3), as re-confirmed by subsequent studies (4, 5). The presence of time-varying amounts of pesticides like diazinon, propanil in aquatic bodies has been reported in many studies. But the presence of glyphosate is unconfirmed, and is unlikely in aquatic bodies that harbor weeds and algae.

In a recent spatio-temporal study (5) pesticide detection occurs in sporadic peaks linked to their application times. Given the short persistence lifetimes of about a day or less for diazinon and propanil at $30^0$-$36^0$ C prevalent in the region studied, Vlahos *et al.* may have overlapped with some local pesticide applications in their *one time point* analysis. We expect that the time-averaged concentration of these pesticides would be quite insignificant, and unlikely to cause *chronic* toxicities that require a persistent insult. Furthermore, the well water is largely consumed as brewed tea or in cooking. Such processing removes steam-volatile substances like diazinon. The admissible *daily* intake (ADI) of, say, diazinon for chronic toxicity for a 70 kg farmer is 0.14 mg/day and the aggregate effect of episodic ingestion can be assessed only if urine and blood data are available.

Jayatilleke *et al.* (6) detected pesticide residues in the urine of CKDu patients as well as in the controls. The proportions of CKDu patients with above-reference values for different pesticide residues greater than 3% were: Chlorpyrifos (10.5%), Carbaryl (10.5%), Naphrhalene (10.5%). No glyphosate was detected within these limits. Vlahos have not reported urine data or other biopsy data.

The pesticides found in the Valhos study are well known for their hepatotoxicity, nephrotoxicity and other effects (8); their nephrotoxicity manifests as glomerular damage. In contrast CKDu shows tubulo-interstitial damage (3) with no signs of hepatotoxicity or other effects typical of these pesticides. CKDu or its rate of progression cannot be correlated with a *single time-point analysis* of pesticide measurement in the wells, or the geographic distribution of CKDu *evolution*, unless we have time-dependent pesticide ingestion data supported by blood and urine analysis.

Furthermore, in Ref. 24 of Vlahos *et al.* they cite the work of Shipley *et al.* (11) who report wells highly contaminated with diazinon, propanil etc., in the upper Mahaweli feeder area south west of the Victoria dam, *but no CKDu has been detected in those areas*. Hence the etiological connection proposed by Vlahos *et al.* between these pesticide contaminations and CKDu do not satisfy even the simplest of Bradford-Hill criteria for causation.

**Hence, in sum we conclude that the Valhos *et al.* data provide no basis for the elucidation of the causes of CKDu or its prevention.**

That the distribution of CKDu may be correlated with the local geomorphology can be seen from the Ginnoruwa area (Figure 1) of Girandurukotte. This is 15-20 km SE of Wilgamuwa studied in (1). Ginnoruwa has three adjacent villages, namely Badulaupura (B), Dolahekanuwa (D), and Sarabhoomiya (S). All three villages were settled in the 1980s and almost all are farmers. However, while most families in B have a CKDu patient, those in D and S have virtually no CKDu patients (4). Figure 1 shows that Badulupura is in rocky high ground where the drinking-water wells are fed from regolith aquifers, while the wells in the CKDu-free low-lying D and S regions are connected with the water table of the paddy fields, as established using isotopic tracer studies (7). The well water in Badulupura (B) is found to contain excessive concentrations of fluoride and magnesium ions (4). Furthermore, excessive fluoride and magnesium concentrations have been found in reverse-osmosis filtration residues in other endemic regions (9), *while no significant agrochemical residues have been detected.* Thus we note that the causation of CKDu proposed by Vlahos *et al*. fails once again to satisfy even the simplest of Bradford-Hill criteria. While any above-threshold presence of pesticides in drinking water even episodically is a public health concern, it is not even correlated with the presence of CKDu.

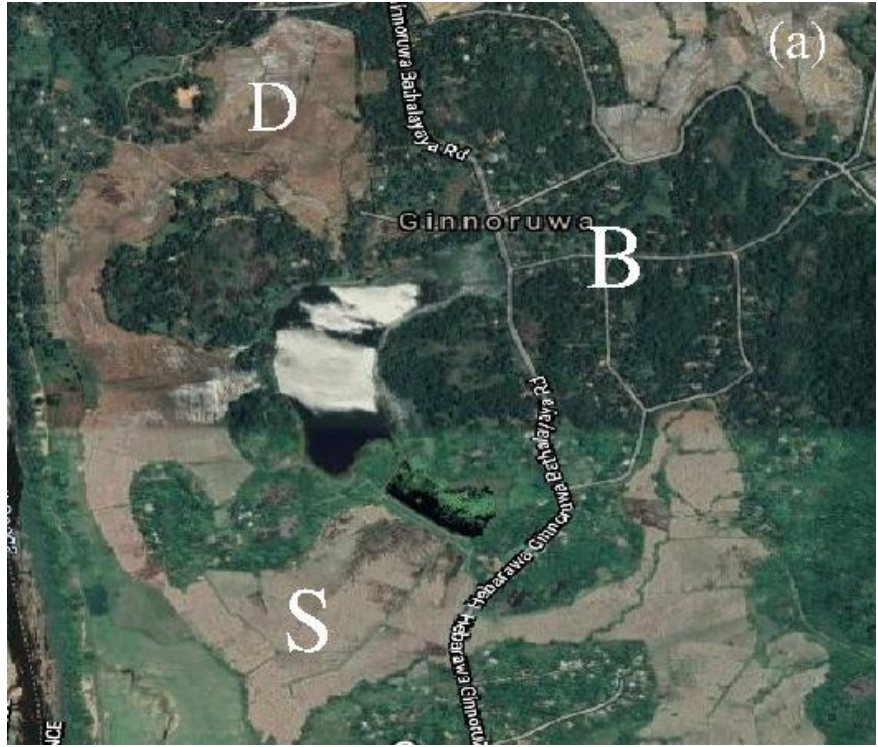

Figure 1. Three adjacent farming villages **Badulupura** (B), **Dolahekaunawa** (D) and **Saramboomiya** (S) in Ginnoruwa (adapted from Google maps). CKDu is *not* endemic to farmers in S and D (4). The wells in the low-lying S, D are connected to agricultural water (6), but the high-ground wells in B are fed by mineral-rich aquifers.

It has been proposed that the combination $Mg^{++}$ and $F^-$ ions may synergistically cause CKDu (10). So, it is natural to ask if the CKDu-associated wells of the Vlahos study (in the Wilgamuwa area) have excess concentrations of magnesium and fluoride ions.

In ancient times, settlers lived in the low-lying valley of the Mahaweli River and depended on it for their water. The 19080s new settlements, away from the river and on higher ground led to the use of untested non-traditional areas, with tube wells and dug wells as sources of drinking water. That they could be contaminated with traces of pesticides, even episodically, points to the need for better controls on the handling of agrochemicals by public health officials.

**References.**